\documentclass[aps,twocolumn,nofootinbib]{revtex4-1}
\usepackage{amsmath,amsthm,amsfonts,amssymb}
\usepackage{graphicx,array,hyperref}

\newcommand{\vectsym}{\boldsymbol}  
\newcommand{\vrho}{\vectsym{\rho}}  

\newcommand {\ket}[1] {\left|{#1}\right\rangle}

\newcommand {\bra}[1] {\langle{#1}|}
\newcommand{\braket}[2]{\langle{#1}|{#2}\rangle}

\newcommand {\cl}{\mathcal}

\newcommand{\val}{\vectsym{\alpha}}

\begin{document}
\title{Realizable receivers for discriminating arbitrary coherent-state waveforms and multi-copy quantum states near the quantum limit}
\author{Ranjith Nair}
\affiliation{Department of Electrical and Computer Engineering, National University of Singapore, 4 Engineering Drive 3, Singapore 117583}
\author{Saikat Guha}
\affiliation{Quantum Information Processing Group,
Raytheon BBN Technologies
Cambridge, MA 02138, USA}
\author{Si-Hui Tan}
\affiliation{Centre for Quantum Technologies, National University of Singapore, 3
Science Drive 2, Singapore 117543}
\date{13 February 2013}
\begin{abstract} 
Coherent states of light, and methods for distinguishing between them, are central to all applications of laser light. We obtain the ultimate quantum limit on the error probability exponent for discriminating among any $M$ multimode coherent-state waveforms via the quantum Chernoff exponent in $M$-ary multi-copy state discrimination. A receiver, i.e., a concrete realization of a quantum measurement, called the Sequential Waveform Nulling (SWN) receiver, is proposed for discriminating an arbitrary coherent-state ensemble using only auxiliary coherent-state fields, beam splitters, and non-number-resolving single photon detectors.   An explicit error probability analysis of the SWN receiver is used to show that it achieves the quantum limit on the error probability exponent, which is shown to be a factor of four greater than the error probability exponent of an ideal heterodyne-detection receiver on the same ensemble. We generalize the philosophy of the SWN receiver, which is itself adapted from some existing coherent-state receivers, and propose a receiver -- the Sequential Testing (ST) receiver--  for discriminating $n$ copies of $M$ pure quantum states from an arbitrary Hilbert space. The ST receiver is shown to achieve the quantum Chernoff exponent in the limit of a large number of copies, and is remarkable in requiring only local operations and classical communication (LOCC) to do so. In particular, it performs adaptive copy-by-copy binary projective measurements. Apart from being of fundamental interest, these results are relevant to communication, sensing, and imaging systems that use laser light and to photonic implementations of quantum information processing protocols in general.
\end{abstract}
\maketitle
The task of optimally discriminating between unknown nonorthogonal quantum states by making appropriate quantum measurements \cite{Hel76,Che00,BC09b,WM10} is a fundamental primitive underlying many quantum information processing tasks, including communication \cite{Hel76}, sensing and metrology \cite{WM10,TEG+08,*Pir11,*Tsa12,*TN12}, and various cryptographic protocols \cite{GRT+02,*Yue09,*MWW09}. The paradigmatic problem of the so-called quantum detection theory \cite{Hel76} is to determine the quantum measurement, specified by a mathematical object known as a positive operator-valued measure (POVM) \cite{NC00,WM10}, that minimizes the average error probability in discriminating a given ensemble of states. The mathematical solution to the problem is known in terms of necessary and sufficient conditions that the optimal POVM must satisfy \cite{YKL75,*BC09}, although for discriminating between more than two states, the explicit solution of these conditions has been obtained only in some specific cases \cite{BKM+97,*EF01,*Bar01,*CH03,*EMV04}. Over the years, the scope of quantum detection theory has been broadened beyond the above framework to ones such as unambiguous state discrimination \cite{Iva87,*Die88,*Per88,*JS95,*PT98,*Che98}, maximum confidence discrimination \cite{CAB+06,*JSD+11}, and to specific scenarios of interest such as  multi-copy state discrimination using local operations and classical communication (LOCC) \cite{BM96,ABB+05,Hay09,HBD+09,CdVM+10,HDB+11}, using a quantum computer with limited entanglement \cite{BCZ12}, and in the asymptotic limit of a large number of copies \cite{NS09,ACM+07,NS11,NS11b}. 
 
The case of multi-copy state discrimination under LOCC and in the asymptotic limit is of particular relevance to this work. For discriminating between $n$ independent and identical copies drawn from one of two density operators, the error probability of the optimal quantum measurement falls off exponentially with $n$  with a characteristic exponent depending on the pair of states, known as the quantum Chernoff exponent \cite{NS09,ACM+07} in analogy with its classical counterpart \cite{Che52}. Although the measurement achieving this optimal scaling behavior is expected to be a joint one over all the $n$ copies, it turns out that, in the case of discriminating two pure states, the scaling and even the exact optimal error probability is obtainable using copy-by-copy measurements with successive measurements depending on previous results, i.e., using adaptive local measurements \cite{ABB+05}. The case of two mixed states is more complicated and is the subject of recent research, although it has been shown using examples of qubits in mixed states that a finite gap exists between the Chernoff exponent and the best exponent achievable using LOCC \cite{Hay09,HBD+09,CdVM+10,HDB+11}. 

The theory of multi-copy state discrimination was recently extended to $M>2$ states in ref.~\cite{NS11}, where it was shown that the error probability scales exponentially with the number of copies with an exponent not larger than the smallest of the pairwise Chernoff exponents (in the sense of refs. \cite{NS09,ACM+07}) between the states of the ensemble. Further, for pure state ensembles, a particular measurement was shown to achieve this exponent, so as to make it the exact $M$-ary Chernoff exponent. In ref.~\cite{NS11b}, this achievability result was extended to the class of linearly independent mixed state ensembles, but has not yet been established in general.

Our work here was originally motivated by the design of concrete \emph{receivers}, i.e., physical realizations of POVM's, for discriminating coherent states of light -- a subject with a long history that remains an active area of research  \cite{Ken73,Dol73,Dol76,CMJ07,SH96,Tak07,TSL06,TSvL+05,TS08,WTS+08,TFF+10,TFF+11,Dol82,CHD+12,Bon93,BFB+11,BFB+12,MUW+12,ITF+12,dSGD12}. Coherent states of light \cite{Gla63} and their random mixtures are the most ubiquitous quantum states of light and their discrimination is central to optical communication \cite{GK95,Agr10} and sensing \cite{WM10,Gas02} with laser light, which is in a coherent state to an excellent approximation. Similar to the situation in multi-copy state discrimination, the optimal error probability of discriminating an ensemble of coherent states decreases exponentially with the average energy, i.e., the average number of photons  in the ensemble \cite{Hel76}, which is the natural resource measure for optical systems. This is also true for receivers that perform the standard direct, homodyne, and heterodyne detections \cite{GK95,Agr10} that correspond to particular POVM's and are realizable in the laboratory \cite{Sha09}. However, the exponent of the optimal receiver allowed by quantum mechanics is in general greater than that of the conventional measurements \cite{Ken73,Dol73,Hel76,Dol82,Bon93}, leaving a gap between the optimal error probability (popularly called the \emph{Helstrom limit}) and the minimum achievable by conventional measurements, viz., homodyne, heterodyne, and direct detection (loosely called \emph{standard quantum limits}).

For discriminating two coherent states, Kennedy proposed a receiver design \cite{Ken73} that is {\em exponentially optimal}, i.e., it achieves the maximum error probability exponent in the high-photon-number regime. Subsequently, Dolinar proposed a more complicated design that exactly achieves the Helstrom limit for discriminating between two coherent state signals \cite{Dol73,Dol76}, which was only recently demonstrated experimentally~\cite{CMJ07}. Sasaki and Hirota conceptualized a `one-shot' receiver that could achieve the Helstrom limit for binary state discrimination without using the fast electro-optic feedback required by the Dolinar receiver ~\cite{SH96}. However, this design required unknown nonlinear-optical transformations, which made it impractical. Takeoka and collaborators showed that an arbitrary binary projective measurement can be performed on arbitrary quantum-optical states using auxiliary coherent fields, linear optics, photon counting and feedback, thereby generalizing Dolinar's receiver beyond coherent states ~\cite{Tak07,TSL06}. Receivers with performance in between the Kennedy and Dolinar receivers were recently proposed \cite{TSvL+05,TS08} for binary coherent state discrimination and experimentally demonstrated \cite{WTS+08,TFF+10,TFF+11} in the low-photon-number regime, where the absolute performance gap between the Kennedy and Dolinar receivers is the largest. For $M>2$, Dolinar proposed a receiver for $M$-ary Pulse Position Modulation (PPM) that is exponentially optimal \cite{Dol82} and was recently demonstrated experimentally \cite{CHD+12}. Bondurant proposed a receiver for the 4-ary Phase Shift Keying (QPSK) constellation that is exponentially optimal \cite{Bon93}. Recently, Becerra and collaborators proposed a feedforward receiver structure for $M$-ary PSK with arbitrary $M$ and demonstrated that it closely approximates the Helstrom limit for $4$-PSK, in a partially simulated experiment without real-time switching of the local oscillator (LO) field \cite{BFB+11}. In work that is to appear \cite{BFB+12}, the Becerra group has implemented their receiver for $4$-PSK with real-time switching of the LO and demonstrated that it beats the standard quantum limit, i.e., the performance of the ideal heterodyne receiver, even without adjusting for detection inefficiency and other realistic limitations for average input photon numbers in the range of $N = 2-15$. Other receiver designs  continue to be proposed and demonstrated for particular coherent-state sets \cite{MUW+12,ITF+12}.

Our contribution in this work is as follows. First, we use the \emph{quantum Chernoff} exponent in $M$-ary multi-copy state discrimination \cite{NS11} to obtain the maximum \emph{error probability} exponent allowed of a coherent-state receiver by quantum mechanics. The Kennedy \cite{Ken73}, Bondurant \cite{Bon93}, and Becerra \cite{BFB+11,BFB+12} receivers rely on a strategy of attempting to null, i.e., displace to the vacuum state, the input state by successively subtracting the fields corresponding to the possible hypotheses. We generalize this strategy to any multimode $M$-ary coherent-state set and propose a receiver for distinguishing between them, which we call the \emph{Sequential Waveform Nulling} (SWN) receiver\footnote{A qualification is in order here. Bondurant, in ref.~\cite{Bon93}, proposed two receivers for the $4$-PSK signal set -- his ``Type I'' and ``Type II'' receivers. The Type I receiver nulls hypotheses in a predetermined sequence, while the Type II receiver nulls hypotheses in an order depending on the times that counts were observed, thereby achieving a slightly improved performance. Similar to the latter receiver, the Becerra \emph{et al.} receiver \cite{BFB+11,BFB+12} nulls, after each detection stage, the most probable hypothesis conditioned on the previous detection data. In contrast, our SWN receiver nulls hypotheses in an arbitrary but predetermined sequence, and is therefore more akin to Bondurant's Type I receiver. However, our claim of asymptotic optimality is expected to carry over, \emph{a fortiori}, to the more optimized strategies of the Becerra \emph{et al.} receivers and Bondurant's Type II receiver. See also ref.~\cite{HDB+11} for a discussion of various strategies to optimize local measurements in the context of binary state discrimination.}. Like its precursors, its operation requires only auxiliary coherent-state generation, beam splitters and single photon detection.  We then compute the error probability of the SWN receiver applied to any state set, and using an upper bound on it, show that the error probability exponent approaches the maximum allowable value, establishing the exponential optimality of the SWN receiver. We also show that the exponent of the multimode heterodyne receiver is smaller in general than that of the SWN receiver by a factor of four. Finally, we adapt the idea of sequential nulling to any $M$-ary ensemble of pure states in an arbitrary Hilbert space, and propose a multi-copy discrimination strategy called the \emph{Sequential Testing} (ST) receiver. We show through an explicit error probability analysis that the ST receiver attains the $M$-ary Chernoff exponent for multi-copy discrimination derived in ref.~\cite{NS11}. Unlike the joint measurement used in ref.~\cite{NS11}, the ST receiver makes only copy-by-copy binary projective measurements, and is thus potentially realizable using current technology. \\ \\

\emph{Coherent-state discrimination --} Before describing the SWN receiver, we lay down some notation and definitions. We are given $M$ quasi-monochromatic complex-valued spatiotemporal coherent-state waveforms $\{\cl{E}_m(\vrho,t)\}_{m=1}^M$, where $\vrho \in \cl{A}$ is the transverse spatial coordinate in the receiver aperture plane $\cl{A}$ and $t \in \cl{T}=[0,T]$ denotes time within the signaling interval $\cl{T}$. The $m$-th waveform corresponds to the $m$-th hypotheses to be discriminated\footnote{For simplicity, we assume the waveforms all have the same polarization.}. The waveforms are given in units of $\sqrt{\rm photons/m^2/s}$, can be completely arbitrary, and correspond to coherent states $\left\{\ket{\val_m}=\ket{\alpha_m^{(1)}} \otimes \cdots \ket{\alpha_m^{(S)}}\right\}_{m=1}^M$  supported on $S \leq M$ orthonormal spatiotemporal modes $\{\phi_s(\vrho,t)\}_{s=1}^S$ that span the waveform space. The $m$-th waveform can then be represented as the point $\val_m \in \mathbb{C}^S$ in an $S$-mode phase space \cite{Sha09}. We define
\begin{align}
E_m  = \int_{\cal{A}} \int_{\cal{T}} \left|\cl{E}_m(\vrho,t)\right|^2 d\vrho\,dt = \parallel{\val_m}\parallel^2
\end{align}
to be the average energy of the $m$-th waveform in photons, and 
\begin{align} \label{Delta}
\Delta_{m,m'}:=&  \int_{\cal{A}} \int_{\cal{T}} \left|\cl{E}_m(\vrho,t)-\cl{E}_{m'}(\vrho,t)\right|^2  d\vrho\,dt \nonumber \\
=& \parallel \val_m - \val_{m'}\parallel^2
\end{align}
to be the energy in the difference of the $m$-th and $m'$-th waveforms. Also define
\begin{align}
\underline{\Delta} := \min_{m,m':m \neq m'} \Delta_{m,m'}.
\end{align}
If the $M$ hypotheses are distributed according to the probability distribution $\{\pi_m\}_{m=1}^M$, the average energy in the waveform ensemble, denoted $N$, is given by
\begin{align}
N= \sum_{m=1}^M \pi_m E_m.
\end{align}

The \emph{error probability exponent} (EPE) $\xi^{\#}$ of a coherent-state receiver $\#$, where $\#$ may denote, e.g., the optimal Helstrom (Hel) receiver, the heterodyne (Het) receiver, or the SWN receiver, is defined as
\begin{align} \label{EPEDef}
\xi^{\#}\left[\{\val_m\}\right]:= - \lim_{N \rightarrow \infty}\frac{1}{N} \ln P_E^{\#}\left[\{\val_m\}^{(N)}\right],
\end{align}
where $P_E^{\#}\left[\{\val_m\}^{(N)}\right]$ is the average error probability of the receiver $\#$ used to discriminate the coherent-state ensemble $\{\ket{\val_m}\}^{(N)}$  consisting of waveforms proportional to the given set of waveforms $\{\val_m\}_{m=1}^M$ but scaled so as to have average energy $N$. Note that this definition is simply the formulation, in the coherent-state context, of the standard notion of error probability exponents in classical digital communication, wherein they function as principal figures of merit for communication systems \cite{PS07}.
\begin{figure*}
\centering
\includegraphics[trim=30mm 48mm 30mm  30mm,clip=true,width=1.4\columnwidth]{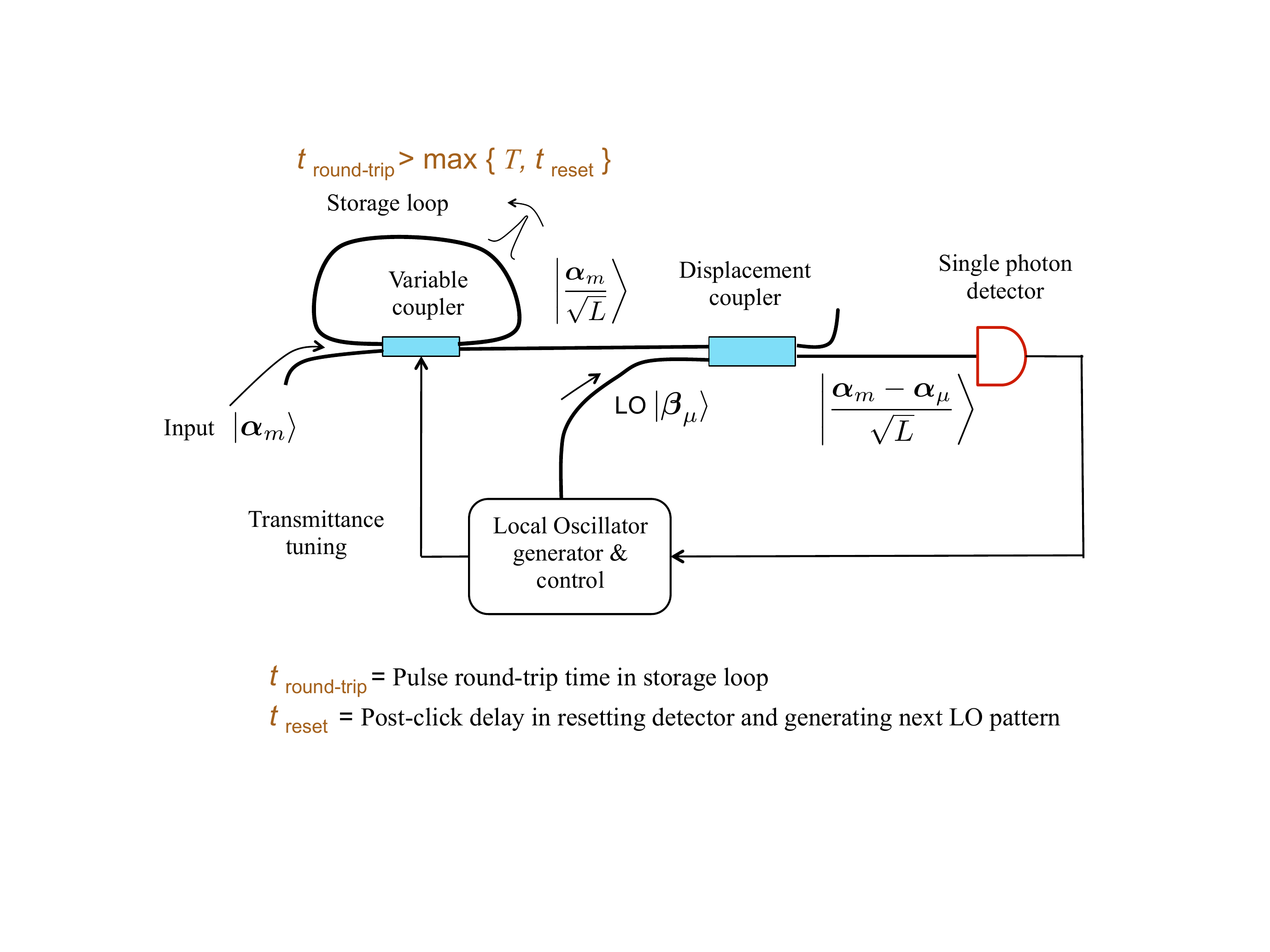}
\caption{Schematic of a possible implementation of the SWN receiver. The input field is effectively split into $L$ equal-amplitude ``slices''. Each slice is displaced by the negative of one of the scaled hypotheses at the displacement coupler and the local oscillator (LO) field pattern is switched to match the next hypothesis for the next slice whenever the single photon detector registers a click. Additional elements necessary to keep the LO amplitude and phase coherent with those of the input are not shown.}
\end{figure*} 

Next, we recall the definition of the $M$-ary quantum Chernoff exponent (QCE) for multi-copy state discrimination from ref.~\cite{NS11}. For any ensemble $\cl{F}=\{\rho_m\}_{m=1}^M$ of states from an arbitrary Hilbert space $\cl{H}$, consider the $n$-copy ensemble $\cl{F}^{\otimes n} = \{\rho_m^{\otimes n}\}_{m=1}^M$. The \emph{quantum Chernoff exponent} (QCE)  $\xi_{\rm QC}\left[\cl{F}\right]$ of $\cl{F}$ is defined as
\begin{align} \label{QCE}
\xi_{\rm QC}\left[\cl{F}\right]:= - \lim_{n \rightarrow \infty}\frac{1}{n} \ln P_E^{\rm Hel}\left[\cl{F}^{\otimes n}\right],
\end{align}
where $P_E^{\rm Hel}\left[\cl{F}^{\otimes n}\right]$ is the average error probability of the \emph{Helstrom} receiver for discriminating the ensemble $\cl{F}^{\otimes n}$. For pure-state ensembles $\cl{F}=\{\ket{\psi_m}\}_{m=1}^M$, it was shown in \cite{NS11} that
\begin{align} \label{PurestateQCE}
\xi_{\rm QC}[\cl{F}]= \min_{m,m':m \neq m'}- \ln \left|\braket{\psi_m}{\psi_{m'}}\right|^2.
\end{align}

For coherent-state ensembles and the Helstrom measurement, these two notions of error probability exponents can be related as follows. The EPE of the Helstrom measurement on a coherent-state ensemble $\{\val_m\}$  is, by definition,
\begin{align}
\xi^{\rm Hel}[\{\val_m\}] :&= - \lim_{N \rightarrow \infty}\frac{1}{N} \ln P_E^{\rm Hel}\left[\{\val_m\}^{(N)}\right] \label{E1} \\
&= - \lim_{n \rightarrow \infty}\frac{1}{n} \ln P_E^{\rm Hel}\left[\{\val_m\}^{(n)}\right] \label{E2}\\
&= - \lim_{n \rightarrow \infty}\frac{1}{n} \ln P_E^{\rm Hel}\left[\otimes^n \{{\val}_m\}^{(1)}\right] \label{E3}\\
&= \xi_{\rm QC}\left[\{{\val}_m\}^{(1)}\right]\\ \label{E4}
\hspace{5mm}= \min_{m,m':m\neq m'}- \ln  &\left|\braket{\val_m^{(1)}}{\val_{m'}^{(1)}}\right|^2 = \underline{\Delta}/N \equiv \kappa.
\end{align}
Here, in eq.~\eqref{E2}, $n$ is restricted to integer values and the equality of \eqref{E1} and \eqref{E2} follows from the existence of the limit of eq.~\eqref{E1}. Eq.~\eqref{E3} follows because a coherent-state ensemble can be split into $n$ identical and independent copies using a unitary beamsplitter transformation and because this action cannot change the error probability of the Helstrom receiver. We are now in the multi-copy discrimination framework and the left-most term in eq.~\eqref{E4} follows from eq.~\eqref{PurestateQCE}, and we have used the coherent-state overlap $\left|\braket{\val_m}{\val_{m'}}\right|^2=e^{- \parallel \val_m-\val_{m'}\parallel^2}$ and defined the constant $\kappa$ that is a  function of the prior probability distribution and the coherent-state ensemble. It is unaffected by scaling all the waveforms of the ensemble by a common factor.

It is interesting to note that the connection made between coherent-state discrimination and multi-copy discrimination via splitting into many identical copies has also recently been exploited to re-derive the Dolinar receiver for discriminating two coherent states \cite{ADP11}. Splitting into many identical copies is also a feature of the receiver proposed in ref.~\cite{dSGD12} for discriminating among any $M$ coherent states at the Helstrom limit. However, this receiver, which builds  on the work of ref.~\cite{BCZ12}, assumes the availability of a quantum computer and the ability to map single-rail photonic qubits onto the Hilbert space of the quantum computer, which is a non-trivial task. \\

\emph{Sequential Waveform Nulling receiver} -- We now describe the operation of the Sequential Waveform Nulling (SWN) receiver (see Fig.~1).
First, the signal field over $\cl{A} \times \cl{T}$ is split into $L$ equal-amplitude portions or \emph{slices} (where $L$ should be as large as possible and at least equal to $(M-2)$ (see below)) that are placed in storage of some kind, e.g., the fiber loop in Fig.~1,  with a view to access the slices sequentially. The receiver now operates as follows.\\\\
\noindent {\bf Sequential Waveform Nulling Receiver}

\begin{enumerate}
\item Initialize the \emph{slice number} $l$ to $l=1.$
\item Initialize the \emph{nulled hypothesis} $\mu$ to $\mu=1$.
\item  While $l \leq L$
\begin{enumerate}
\item Displace the $l$-th slice of the input field by the field $-\frac {\cl{E}_{\mu}(\vrho,t)}{\sqrt{L}}$ and direct-detect the output field in $\cl{A} \times \cl{T}$ on a single photon detector.
\item If the detector clicks, set $\mu := \mu+1.$
\item $l := l+1.$
\end{enumerate}

\item Set the receiver's decision $\hat{m} := \mu$.
\end{enumerate}

Before analyzing the receiver's performance, we comment on why the $L$-fold amplitude slicing of the signal is useful. Indeed, a natural way to generalize the Kennedy \cite{Ken73} and Bondurant \cite{Bon93} receivers is to stipulate that one displace the input field at $(\vrho,t)$ by the negative of $\cl{E}_{\mu}(\vrho,t)$, where $(\mu-1)$ equals the number of clicks observed in $[0,t]$ (see also the Conditional Pulse Nulling (CPN) receiver of ref.~\cite{GHT11} that follows the same procedure for hypotheses in a single spatial mode). While doing so achieves the optimum exponent for waveforms in a single temporal mode, there are ternary discrimination problems with signals in two temporal modes for which, at the very least, the order in which the hypotheses are nulled matters and complicates the performance analysis \cite{NGT12-unp}.  This `continuous-time' version of the sequential nulling strategy will not be discussed further here -- we concentrate on the amplitude-sliced SWN receiver alone. As we show below, the slicing strategy leads to the SWN receiver achieving the optimal exponent for an \emph{arbitrary multimode} set of hypotheses in the limit of large $L$.

Slicing  also serves a practical purpose. The limitations on the speed of electro-optic switching of the LO waveform, as well as the finite dead time of single photon detectors such as APDs following the detection of a photon mean that detection cannot continue immediately after a detector click. If the next slice is held in storage until the LO waveform and detector are reset, we need not lose the portion of the input state in this reset period. Indeed, such a splitting strategy was already employed, albeit with a different architecture than Fig.~1, in the Becerra \emph{et al.} experiment of ref.~\cite{BFB+11} (Our slices are referred to therein as ``stages'').  In their more recent experiment in ref.~\cite{BFB+12}, amplitude slicing was replaced by the theoretically equivalent strategy -- for the flat-top input pulses used in \cite{BFB+12} -- of slicing the input pulse temporally into  slices of equal duration.

The error probability analysis of the SWN receiver requires only the semiclassical theory of photodetection \cite{GK95,Sha09}, and is carried out in the supplementary material, where the exact error probability is derived. The following upper bound on the average error probability $P^{\rm SWN}_E\left[\{\val_m\}\right]$ is also obtained therein --
\begin{align} 
P^{\rm SWN}_E\left[\{\val_m\}\right] &= \sum_{m=1}^M \pi_m P[\,E \,|\,m] \nonumber \\ 
& \leq  \sum_{m=1}^M \pi_m\sum_{K=0}^{m-2}{L \choose K} e^{-\kappa\left(1- \frac{K}{L}\right)N},\label{PeUB}
\end{align}
where $P[\,E \,|\,m]$ is the conditional probability of error given that the true state was $\ket{\val_m}$. A lower bound on the EPE $\xi^{\rm SWN}\left[\{\val_m\}\right]$ follows on substituting the right-hand side of eq.~\eqref{PeUB}
into that of eq.~\eqref{EPEDef}. A calculation shows that the latter bound depends only on the slowest decaying exponential term of  \eqref{PeUB}, so that we have
\begin{align} \label{xiSWN}
\xi^{\rm SWN}\left[\{\val_m\}\right] &\geq \left(1-\frac{M-2}{L}\right) \kappa  \nonumber \\
&= \left(1-\frac{M-2}{L}\right) \xi^{\rm Hel}\left[\{\val_m\}\right].
\end{align}
Because we must have $\xi^{\rm SWN}\left[\{\val_m\}\right] \leq \xi^{\rm Hel}\left[\{\val_m\}\right]$ by definition of the Helstrom receiver, we conclude that the EPE of the SWN receiver is at most a factor of $(1-(M-2)/L)$ away from the Helstrom receiver. If $L<(M-2)$, we lack slices to see enough clicks to ever declare the $M$-th (and perhaps more hypotheses, depending on how much $L$ falls short of $(M-2)$) hypothesis, and $\xi^{\rm SWN}[\{\val_m\}]=0$ (see the derivation of eq.~\eqref{PeUB} in the supplementary material). On the other hand, in the limit of $L \rightarrow \infty$, the two exponents must be identical, establishing the optimality of the SWN receiver exponent in this limit. For a given $M$, $L$ need not be very large for the EPE to be close to optimal, as seen in the example below.

 \begin{figure}
\centering
\includegraphics[trim=80mm 70mm 45mm  52.5mm,clip=true,width=1.4\columnwidth]{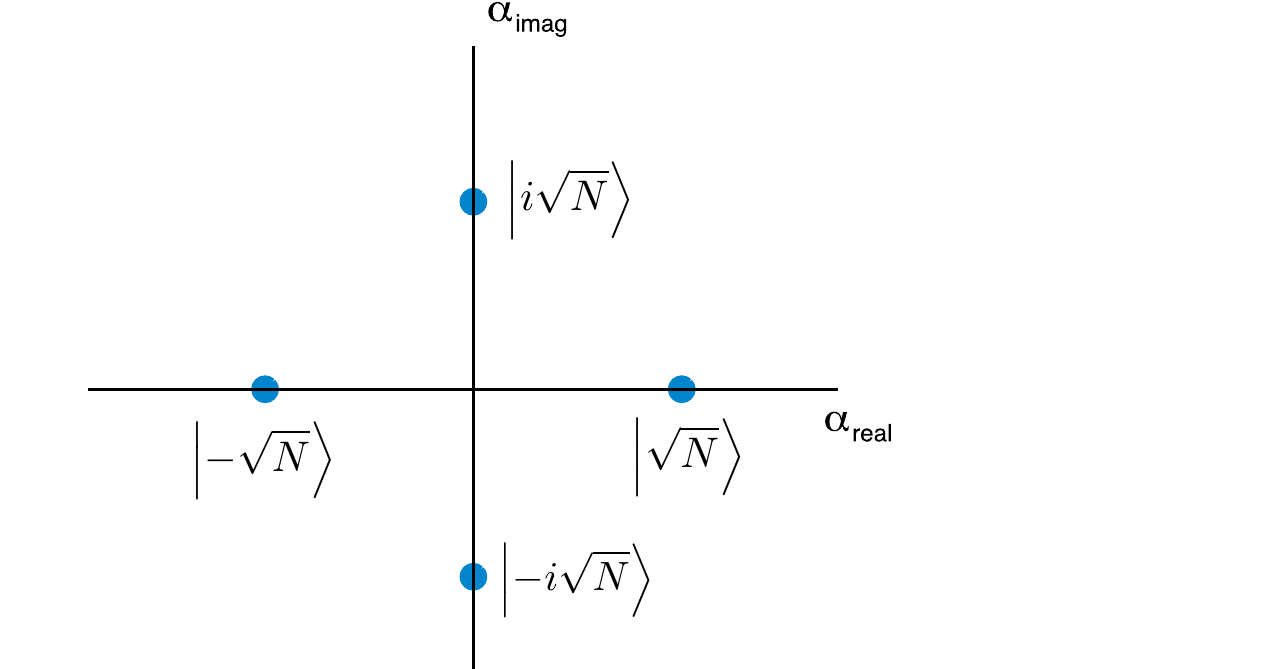}
\caption{Coherent-state constellation for quadrature phase-shift keying.}
\end{figure}
\begin{figure*}
\centering
\includegraphics[trim=23mm 79mm 29mm  86mm,clip=true,width=1.4\columnwidth]{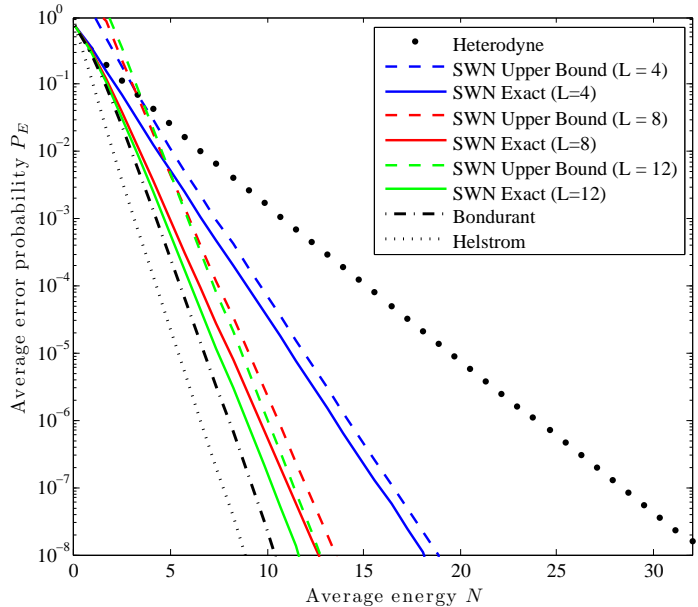}
\caption{Error probability behavior of various receivers for an ensemble of equally likely quadrature phase-shift-keyed coherent states as a function of the average ensemble energy. The colored dashed lines are, for $L=4,8$ and 12, the error probability bound eq.~\eqref{PeUB}, with the corresponding solid lines being the exact error probability (see supplementary material) of the SWN receiver. See main text for discussion.}
\end{figure*}

\begin{figure*}
\centering
\includegraphics[trim=23mm 79mm 29mm  86mm,clip=true,width=1.4\columnwidth]{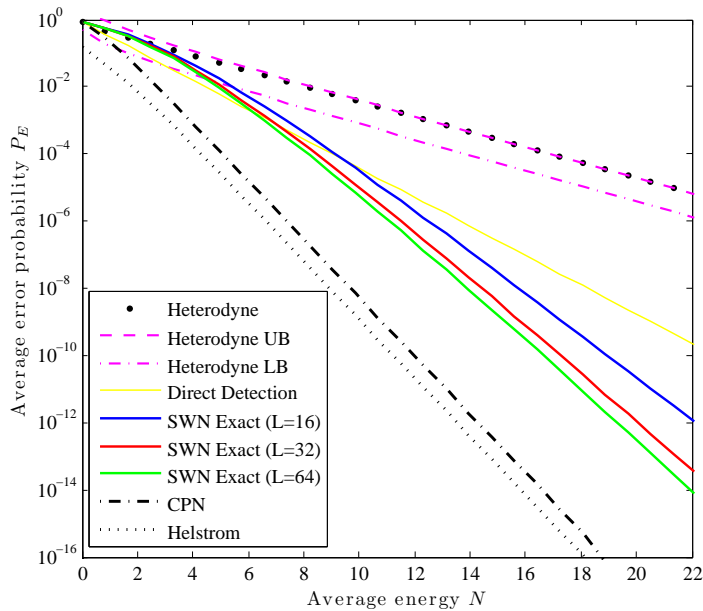}
\caption{Error probability behavior of various receivers for an ensemble of equally likely PPM states with $M=6$ pulses as a function of the average ensemble energy. The heterodyne upper bound is the union bound. See main text for discussion.}
\end{figure*}

\emph{Comparison with heterodyne exponent} -- For a coherent-state ensemble $\{\ket{\val_m}\}$ supported on $S$ modes, we may in principle heterodyne each of the $S$ modes to get, conditional on the state $\ket{\val_m}$, an observation in $\mathbb{C}^S$ which equals $\val_m$ with added Gaussian noise of variance $1/2$ in each of $2S$ orthogonal quadratures of the $S$ modes. For this essentially classical situation involving a fixed measurement, it is known that the $M$-ary Chernoff exponent equals the worst-case binary Chernoff exponent between all pairs $m$ and $m'$ of the hypotheses \cite{NS11b,Sal99}. This latter quantity is known to equal $d^2/8\sigma^2$ for two one-dimensional Gaussian distributions with the same variance $\sigma^2$ and means separated by $d$ (see, e.g., Ch.~2 of \cite{Van01}), which leads in the heterodyne case to
\begin{align}
\xi^{\rm Het}\left[\{\val_m\}\right]= \min_{m,m':m\neq m'} \frac{\parallel \val_m - \val_{m'}\parallel^2}{4N}= \frac{\kappa}{4},
\end{align}
so that the heterodyne EPE is a factor of $4$ worse than the Helstrom EPE regardless of the ensemble $\{\ket{\val_m}\}$.

The performance of various receivers on the single-mode quadrature phase-shift keyed signal set of Fig.~2 is compared in Fig.~3, with the hypotheses assumed \emph{a priori} equally likely. The exponential scaling of all the receivers is apparent in the straight-line dropoff of the error probability already evident at rather low photon numbers of $N \sim 5$. The Helstrom measurement (which in this case implements the square-root measurement \cite{KOS+99}) and the Bondurant receiver \cite{Bon93}, which is equivalent to the $L=\infty$ version of the SWN receiver, have the same EPE. The exact error probability of the SWN receiver and the upper bound of eq.~\eqref{PeUB}, are shown as the colored solid and dashed lines respectively in Fig.~3 for $L=4, 8$ and 12. Note that the difference in slopes of the $L=12$ error probability and the Helstrom error probability is slight. Finally, the heterodyne receiver error probability is seen to have a markedly slower dropoff with $N$ than the other receivers.\\\

The performance of various receivers on the PPM signal set for $M=6$ pulses with the hypotheses assumed \emph{a priori} equally likely is shown in Fig.~4. The exact error for this signal set using direct detection and the CPN receiver have been derived in \cite{GHT01}. The CPN receiver approaches the same EPE as the Helstrom measurement. Although the CPN outperforms the SWN at all $N$ and $L$, the difference in slopes of the $L=64$ SWN and the Helstrom error probabilities is slight. This indicates that SWN is capable of almost approaching the same EPE as the Helstrom measurement. In Fig.~4, we also show the exact error probability for heterodyne detection along with the union upper bound and the lower bound of eq.~(3.7.3) of  \cite{VO09}. The heterodyne receiver error exponent performs a lot worse than the SWN receivers, the CPN receiver and direct detection.

\emph{Sequential Testing receiver} -- The philosophy of the SWN receiver may be applied back to the scenario of discriminating multiple copies of pure states of ref.~\cite{NS11}. Abstractly, the nulling process implements a unitary transformation that maps the state $\ket{\val_{\mu}}$ corresponding to the nulled hypothesis to a standard state (namely, the multimode vacuum state $\ket{\mathbf{0}}$). Detection using a single photon detector corresponds to measuring the two-outcome POVM $\{\ket{\mathbf{0}}\bra{\mathbf{0}}, \mathbb{I}- \ket{\mathbf{0}}\bra{\mathbf{0}}\}$. Taken together, these two steps yield the same statistics, for any input state, as a measurement of the POVM $\{\ket{\val_{\mu}}\bra{\val_{\mu}}, \mathbb{I}- \ket{\val_{\mu}}\bra{\val_{\mu}}\}$. Conditioned on the outcome, further similar measurements are performed on independent copies of the input state. Note that we do not use the post-measurement state of any of these sub-measurements, even if it may be available, and therefore, the measurements may be destructive (as is the case for the SWN receiver). 

Consider a pure-state ensemble $\cl{F}= \{\ket{\psi_1}, \ldots \ket{\psi_M}\}$ of states on an arbitrary Hilbert space $\cl{H}$ with prior probabilities $\{\pi_m\}_{m=1}^M$. For each $\ket{\psi_m} \in \cl{F}$,  define a binary projective POVM on $\cl{H}$ with elements $\Pi_m = \ket{\psi_m}\bra{\psi_m}$ and $\Pi_m^{\perp}= 1-\Pi_m$. The unknown input state $\ket{\psi_m}$ enters the receiver in the $n$-copy form  $\otimes_{l=1}^n \ket{\psi_m}_l$, where $1\leq l \leq n$ denotes the copy  index. The Sequential Testing (ST) receiver operates as follows. \\\\
\noindent {\bf Sequential Testing Receiver}

\begin{enumerate}
\item Initialize the \emph{copy index} $l$ to $l=1.$
\item Initialize the \emph{current hypothesis} $\mu$ to $\mu=1$.
\item  While $l \leq n$
\begin{enumerate}
\item Measure $\{\Pi_\mu,\Pi_\mu^\perp\}$ on $\ket{\psi_m}_l$.
\item If the $\Pi_\mu^\perp$ outcome is obtained, set $\mu := \mu+1.$
\item $l := l+1.$
\end{enumerate}

\item Set the estimated hypothesis $\hat{m} := \mu$.
\end{enumerate}

Analogous to the quantum Chernoff exponent defined in eq.~\eqref{QCE}, we may define the multi-copy error exponent $\xi^{\rm ST}[\cl{F}]$ of the ST receiver as
\begin{align} \label{xiSTDef}
\xi^{\rm ST}\left[\cl{F}\right]:= - \lim_{n \rightarrow \infty}\frac{1}{n} \ln P_E^{\rm ST}\left[\cl{F}^{\otimes n}\right],
\end{align}
where $P_E^{\rm ST}\left[\cl{F}^{\otimes n}\right]$ is the average error probability of the ST receiver for discriminating the ensemble $\cl{F}^{\otimes n}$. In the supplementary material, via an analysis paralleling that for the SWN receiver, we obtain the upper bound 
\begin{align} \label{STPeUB}
P_E^{\rm ST}\left[\cl{F}^{\otimes n}\right] \leq \sum_{m=1}^M \pi_m \sum_{K=0}^{m-2}{n \choose K}  F_{\rm max}^{(n-K)},
\end{align}
where $F_{\rm max}$ is the maximum pairwise fidelity between the states of the ensemble --
\begin{align}
F_{\rm max}= \max_{m,m':m \neq m'}\left|\braket{\psi_m}{\psi_{m'}}\right|^2.
\end{align}
We further show that this bound implies that $\xi^{\rm ST} \geq \xi_{\rm QC}$. Since, by definition, $\xi^{\rm ST} \leq \xi_{\rm QC}$, we must have
\begin{align}
\xi^{\rm ST} = \xi_{\rm QC},
\end{align}
so that the ST receiver achieves the quantum Chernoff exponent in the asymptotic limit of many copies.

\emph{Discussion} -- We have described the operation of two practical receivers, one for general $M$-ary coherent-state discrimination, and the other for discriminating among multiple copies of any ensemble of pure quantum states. The SWN receiver uses only beam splitters, the ability to engineer arbitrary coherent-state waveforms, as provided by arbitrary optical waveform generators (AOWG's) and spatial light modulators (SLM's), and single photon detection, as provided by superconducting nanowire single-photon photodetectors (SNSPD's), transition edge sensor single photon detectors (TES's), or even avalanche photodiodes (APD's).  Remarkably, this limited toolbox of operations achieves the optimal error exponent allowed by quantum mechanics for discriminating coherent-state waveforms.  The SWN receiver does not rely on nonlinear optics processes, which simplifies its implementation for optical communication over unamplified channels such as deep-space and satellite communication links, for metrology and imaging using laser light, and for coherent-state cryptographic schemes. In further research, apart from exploring the various potential applications, it would be useful to study the degrading effect of errors in the displacement process arising from imperfect control over the phase and amplitude of the LO waveforms.

Similar to the SWN receiver, the ST receiver has many attractive features. Besides attaining the $M$-ary quantum Chernoff exponent for pure states (the same strategy was known to do so for $M=2$ \cite{ABB+05}), it is remarkable that it does so using only local operations and classical communication (LOCC), the previously known receiver attaining the Chernoff exponent being a joint measurement in the Gram-Schmidt basis defined by the multi-copy ensemble \cite{NS11}. Furthermore, the measurements on each copy of the input are binary projective measurements and can be destructive -- a feature that is attractive for photonic implementations. Such measurements are readily made on polarization-encoded photonic qubits, as e.g., in the experiments of ref.~\cite{HBD+09}. As another example, the required measurements on multi-copy ensembles of displaced squeezed states \cite{Yue76} can be made using displacement operations, squeezing, and on-off photodetection. Finally, the work of ref.~\cite{TSL06} implies that any multimode binary projective measurement on any given multimode optical state of light and its orthogonal complement can be performed using auxiliary coherent-state local oscillators, single photon detection, and classical feedforward. This will enable an implementation of the ST receiver for general ensembles of optical states, including nonclassical states of light. We hope that our work stimulates further theoretical study of methods for $M$-ary multi-copy state discrimination, as well as experimental implementations of the receivers presented here.\\

The authors thank  A.~Chia, M.~Nussbaum, M.~Tsang, H. M.~Wiseman, and B. J.~Yen for helpful discussions. R.N. is supported by the Singapore National Research Foundation under NRF Grant No. NRF-NRFF2011- 07. S. G. was supported by the DARPA Information in a Photon (InPho) program under contract number HR0011-10-C-0162. S.-H. T. was supported by the Data Storage Institute, Singapore.

\bibliography{STSWN8}
\bibliographystyle{apsrev}
\onecolumngrid
\newpage
\appendix*
\section*{Supplementary Material}
\subsection{SWN Receiver Performance}
In this appendix, we derive an expression for the error probability of the SWN receiver for an arbitrary coherent-state ensemble and also prove the upper bound eq.~\eqref{PeUB} on it. 

From the way the receiver operates, it is apparent that when the $m$-th hypothesis is true, we cannot get more than $m-1$ total clicks over the $L$ slices. Further, if  $m-1$ clicks are observed, we declare correctly that hypothesis $m$ is true.  We thus have for the conditional probability of error given that hypothesis $m$ is true --
\begin{align}
P^{\rm{SWN}}[E|m] &= \textrm{Pr}\,[ \textrm{Fewer than}\;m-1 \;\textrm{clicks are observed}\,|\,m\,] \nonumber\\ &=\sum_{K=0}^{m-2}\textrm{Pr}\,[ K \;\textrm{clicks are observed}\,|\,m\,],
\end{align}
where the $m=1$ case may be included by agreeing that sums in which the starting value of the summation index exceeds the ending value are zero. For $K>0$, the summand may be written as follows -- the $K=0$ case is dealt with later. Define a length-$K$ vector $\mathbf{l} = (l_1, \ldots,l_K)$ whose $k$-th component $l_k$ is the slice number in the detection of which the $k$-th click occured. The possible instances of $\mathbf{l}$ are the 
increasing sequences of $K$ integers chosen from $\{1, \ldots, L\}$, and are thus ${L \choose K}$ in number.  When the nulled hypothesis is $\mu <m$, the average number of photons incident on the detector in one slice is $\Delta_{\mu,m}/L$. We may then write, using the Poisson nature of the count in each slice together with the conditional statistical independence of photodetection in successive slices:
\begin{align}
&\textrm{Pr}\,[ K \;\textrm{clicks are observed}\,|\,m\,]  \nonumber \\ \label{PrKclicks|m}
=& \sum_{{\rm allowed}\; \mathbf{l}} \exp{\left\{-\Delta_{1,m}\frac{\left(l_1-1\right)}{L}\right\}}\left(1-\exp{\left\{-\frac{\Delta_{1,m}}{L}\right\}} \right) \times \nonumber \\
&\exp{\left\{-\Delta_{2,m}\frac{(l_2-l_1-1)}{L}\right\}}\left(1-\exp{\left\{-\frac{\Delta_{2,m}}{L}\right\}}\right) \cdots \nonumber \\ 
&  \cdots\exp{\left\{-\Delta_{2,m}\frac{(l_K-l_{K-1}-1)}{L}\right\}} \left(1-\exp{\left\{-\frac{\Delta_{K,m}}{L}\right\}}\right)\  \nonumber \\
& \times \exp{\left\{-\Delta_{K+1,m}\frac{(L-l_K)}{L}\right\}},
\end{align}
where factors of the form $\exp\{\cdot\}$ are probabilities that no clicks are obtained in the intervals between the click locations indicated by $\mathbf{l}$ while factors of the form $(1-\exp\{\cdot\})$ are probabilites of obtaining a click in the click locations. We may bound the above expression as
\begin{align}
&\textrm{Pr}\,[ K \;\textrm{clicks are observed}\,|\,m\,]  \nonumber \\
& \leq \sum_{{\rm allowed}\; \mathbf{l}} \exp{\left\{-\Delta_{1,m}\frac{(l_1-1)}{L}\right\}} \exp{\left\{-\Delta_{2,m}\frac{(l_2-l_1-1)}{L}\right\}}\nonumber \\
&\cdots \exp{\left\{-\Delta_{K+1,m}\frac{(L-l_K)}{L}\right\}} \\
& \leq \sum_{{\rm allowed}\; \mathbf{l}} \exp{\left\{-\underline{\Delta}\frac{(l_1-1)}{L}\right\}} \exp{\left\{-\underline{\Delta}\frac{(l_2-l_1-1)}{L}\right\}} \nonumber \\
&\cdots \exp{\left\{-\underline{\Delta}\frac{(L-l_K)}{L}\right\}} \\
& = {L \choose K} \exp{\left\{-\underline{\Delta} \frac{(L-K)}{L}\right\}} \label{PrKclicksUB|m}.
\end{align}
For $K=0$ and for $m>1$, we have $\textrm{Pr}\,[ K \;\textrm{clicks are observed}\,|\,m\,] = \exp(-\Delta_{1,m}) \leq  \exp(-\underline{\Delta})$. The upper bound of eq.~\eqref{PrKclicksUB|m} is therefore also valid in this case. For $m=1$, we have  $P^{\rm{SWN}}[E|m=1]=0$. Therefore, keeping to the summation convention adopted above, we may write, for all values of $m$,
\begin{align}
 P^{\rm{SWN}}[E|m] 
\leq \sum_{K=0}^{m-2}{L \choose K} e^{-\underline{\Delta}\, \frac{(L-K)}{L}},
\end{align}
so that the total error probability of the SWN receiver is 
\begin{align} \label{PE}
P^{\rm{SWN}}_E\left[\{\val_m\}\right]= \sum_{m=1}^M \pi_m P[\,E \,|\,m] \nonumber \\ 
\leq  \sum_{m=1}^M \pi_m\sum_{K=0}^{m-2}{L \choose K} e^{-\underline{\Delta}\, \frac{(L-K)}{L}},
\end{align}
which is eq.~\eqref{PeUB} of the main text. A lower bound on the EPE $\xi^{\rm{SWN}}\left[\{\val_m\}\right]$ of the SWN receiver can be obtained by inserting the right-hand side of \eqref{PE} into \eqref{EPEDef}. Only the term that decays the slowest with $\underline{\Delta}$ (or equivalently, with $N$) survives in the limit of $N \rightarrow \infty$, so that
\begin{align} \label{XiBRbound}
\xi^{\rm{SWN}}\left[\{\val_m\}\right] \geq \frac{\underline{\Delta}}{N}\left(1-\frac{M-2}{L}\right) = \kappa\left(1-\frac{M-2}{L}\right),
\end{align}
which is eq.~\eqref{xiSWN} of the main text. As $L$ is increased, this lower bound approaches $\kappa$ arbitrarily closely.
\subsection{ST Receiver Performance}
In this section, we derive the error probability of the ST receiver, and via an upper bound on this probability, we establish that it attains the $M$-ary quantum Chernoff exponent in the asymptotic limit.

The performance analysis of the ST receiver follows largely the same lines as that of the SWN receiver.  Because making the $\{\Pi_m,\Pi_m^\perp\}$ measurement on $\ket{\psi_m}$ can never lead to a `$\perp$' outcome, we have
\begin{align}
P^{\rm{ST}}[E|m] &= \textrm{Pr}\,[ \textrm{Fewer than}\;(m-1) \;{\textrm `}\perp\textrm{' outcomes}\,|\,m\,] \\ &=\sum_{K=0}^{m-2}\textrm{Pr}\,[ K \;\;{\textrm `}\perp\textrm{' outcomes}\,|\,m\,].
\end{align}
 As before, for each $K>0$, we define a length-$K$ vector $\mathbf{l} = (l_1, \ldots,l_K)$ whose $k$-th component $l_k$ is the copy number in the detection of which the $k$-th `$\perp$' outcome occured. The possible instances of $\mathbf{l}$ are the increasing sequences of $K$ integers chosen from $\{1, \ldots, n\}$, and are thus ${n \choose K}$ in number. We may then write
\begin{align}
&\textrm{Pr}\,[ K \;\;{\textrm `}\perp\textrm{' outcomes}\,|\,m\,] = \nonumber \\ \label{PrKclicks|m}
=& \sum_{{\rm allowed}\; \mathbf{l}} \left\vert\braket{\psi_1}{\psi_m}\right\vert^{2\left(l_1-1\right)} \cdot\,\left[1-\left\vert\braket{\psi_1}{\psi_m}\right\vert^2\right]\cdot\,\left\vert\braket{\psi_2}{\psi_m}\right\vert^{2\left(l_2-l_1-1\right)}\cdot\left[1-\left\vert\braket{\psi_2}{\psi_m}\right\vert^2\right] \cdots \nonumber \\
&\;\;\;\;\;\;\;\;\;\;\;\indent \indent \indent \indent \cdots \left[1-\left\vert\braket{\psi_K}{\psi_m}\right\vert^2\right] \cdot \, \left\vert\braket{\psi_{K+1}}{\psi_m}\right\vert^{2(n-l_K)}  \\
& \leq \sum_{{\rm allowed}\; \mathbf{l}} F_{\rm max}^{\left(l_1-1\right)} \cdot\,F_{\rm max}^{(l_2-l_1-1)}\cdots \, F_{\rm max}^{(n-l_K)}  \\
& = \sum_{{\rm allowed}\; \mathbf{l}} F_{\rm max}^{(n-K)} = {n \choose K}  F_{\rm max}^{(n-K)}, \label{Fmaxbound}
\end{align}
where
\begin{align}
F_{\rm max} = \max_{m,m':m \neq m'} \left|\braket{\psi_m}{\psi_{m'}}\right|^2
\end{align}
and \eqref{Fmaxbound} happens to also be valid for $K=0$. The average error probability of the ST receiver can then be bounded as 
\begin{align} \label{PESTR}
P^{\rm{ST}}_E\left[\{\ket{\psi_m}^{\otimes n}\}\right]= \sum_{m=1}^M \pi_m P[\,E \,|\,m] \leq  \sum_{m=1}^M \pi_m\sum_{K=0}^{m-2}{n \choose K} F_{\rm max}^{(n-K)}.
\end{align}
 A lower bound on the Chernoff error exponent  $\xi^{\rm{ST}}$ of the ST receiver can be obtained by inserting the right-hand side of \eqref{PESTR} into \eqref{xiSTDef}. Factoring out the lowest power of $F_{\rm max}$ on the right-hand side, namely $F_{\rm max}^{n-M+2}$, we have
\begin{align} \label{EA1}
-\frac{1}{n} \ln P^{\rm{ST}}_E\left[\{\ket{\psi_m}^{\otimes n}\}\right] &\geq -\frac{n-M+2}{n}\;\ln F_{\rm max} - \frac{1}{n} \ln \left[ \sum_{m=1}^M \pi_m\sum_{K=0}^{m-2}{n \choose K} F_{\rm max}^{(M-K-2)}\right]
\end{align}
Since $\pi_m \leq 1$, ${n \choose K} \leq n^K$, and $F_{\rm max}<1$, we may bound the argument of the logarithm in the second term as
\begin{align}
 \sum_{m=1}^M \pi_m\sum_{K=0}^{m-2}{n \choose K} F_{\rm max}^{(M-K-2)} &\leq  M^2 n^{M-2}.
\end{align}
Substituting this back into (\ref{EA1}) gives
\begin{align}
-\frac{1}{n} \ln P^{\rm{ST}}_E\left[\{\ket{\psi_m}^{\otimes n}\}\right] &\geq -\frac{n-M+2}{n}\;\ln F_{\rm max} - \frac{\ln M^2 }{n} - (M-2)\frac{\ln n}{n}
\end{align}
Taking the limit of $n \rightarrow \infty$, we have
\begin{align} \label{XiBRbound}
\xi^{\rm{ST}}\left[\{\ket{\psi_m}\}\right] \geq - \ln F_{\rm max}= \xi_{\rm QC}\left[\{\ket{\psi_m}\}\right].
\end{align}
Since $\xi_{\rm QC}\left[\{\ket{\psi_m}\}\right]$ is the maximum exponent allowed of any receiver, we must have
\begin{align}
\xi^{\rm ST}\left[\{\ket{\psi_m}\}\right] = \xi_{\rm QC}\left[\{\ket{\psi_m}\}\right].
\end{align}


\end{document}